\documentclass[12pt,english]{article}
\usepackage[T1]{fontenc}
\usepackage[latin9]{inputenc}
\usepackage{amsmath}
\usepackage{amssymb}
\usepackage{graphicx}

\makeatletter

\date{}

\newcommand{\Pibar}{\overline{\Pi}}
\newcommand{\abar}{\bar{a}}
\newcommand{\bbar}{\bar{b}}
\newcommand{\xibar}{\bar{\xi}}
\newcommand{\taubar}{\bar{\tau}}
\newcommand{\Omegabar}{\bar{\Omega}}
\newcommand{\Xbar}{\overline{X}}
\newcommand{\Ybar}{\overline{Y}}
\newcommand{\Zbar}{\overline{Z}}

\newcommand{\Wbar}{\overline{W}}
\newcommand{\Fbar}{\overline{F}}

\usepackage{babel}

\makeatother

\usepackage{babel}
\begin{document}

\title{Numerical Evaluation of Accelerated-Assisted Entanglement Harvesting}

\author{Andrew Brainerd\thanks{ab3418@columbia.edu} and Brian Greene\thanks{greene@phys.columbia.edu}\\
\emph{Institute of Strings, Cosmology, and Astroparticle Physics}\\
\emph{Center for Theoretical Physics}\\
\emph{Columbia University, New York, NY 10027, USA}}
\maketitle

\begin{abstract}
We consider acceleration-assisted entanglement harvesting as evidenced in correlations between two accelerating Unruh detectors coupled to a scalar field. We elaborate on earlier studies, which in a stationary phase approximation calculated the entanglement dependence on two parameters $c_1 = \kappa L$, and $c_2 = \kappa \Omega \sigma^2$, where $\kappa$ describes the detector's acceleration, $L$ their separation and $\Omega$ the energy splitting in a pair of two state Unruh detectors.  Here, we go beyond the stationary phase approximation by performing a numerical calculation of entanglement harvesting, allowing us to present the dependence on  $c_3 = \sigma \Omega$, where $\sigma$ denotes the half width of a Gaussian window function specifying the field-detector interaction, and show agreement with earlier work the large $c_3$ limit. 
\end{abstract}
\global\long\def\Omegabar{\bar{\Omega}}

\global\long\def\Xbar{\overline{X}}

\global\long\def\Ybar{\overline{Y}}

\global\long\def\Zbar{\overline{Z}}

\global\long\def\xibar{\bar{\xi}}
 \global\long\def\taubar{\overline{\tau}}
 \global\long\def\Wbar{\overline{W}}
 \global\long\def\abar{\overline{a}}
 \global\long\def\bbar{\overline{b}}
 \global\long\def\Fbar{\overline{F}}
 \global\long\def\Pibar{\overline{\Pi}}

\section{Introduction}
\label{previous-calculations}

Over the last few years, there has been investigation into "entanglement harvesting" \cite{Salton} \cite{Nambu} \cite{Martin-Martinez} \cite{VerSteeg} \cite{Reznik} \cite{Reznik2}: a phenomenon, most easily realized in models containing a scalar field coupled to multiple (usually two) separated Unruh-DeWitt
detectors, in which, for certain choices of the detectors'
worldlines, they can become quantum entangled.  In a sense, the entangled nature of the vacuum state of a scalar field can be transfered to detectors with appropriate interactions and executing suitable motions. 

Entanglement harvesting is a beautiful illustration of how the infectious nature of entanglement allows interactions to readily spread this iconic quantum characteristic. Moreover, entanglement harvesting provides a simple laboratory to study how the degree to which two objects -- in our case, two Unruh-DeWitt detectors-- become entangled depends on detailed physical features including the accelerations of the detectors, the mass gap of each detector, and the distance between them.

In Salton, et al. \cite{Salton}, the authors used the by now standard measure of entanglement, "negativity"  (reviewed briefly below) to quantify the entanglement between two accelerating Unruh-DeWitt detectors. Using repeated stationary phase approximations, the authors found the region in the space of coeficients $(c_1, c_2)$ for which the Unruh-DeWitt detectors would become entangled, where $c_1 = \kappa L$, and $c_2 = \kappa \Omega \sigma^2$, where $\kappa$ describes the relative acceleration, $L$ the separation and $\Omega$ the energy splitting in a pair of two state Unruh detectors.  Of particular note, in the stationary phase approximation invoked, the parameter 
$c_3 = \sigma \Omega$, with $\sigma$ denoting the half width of a Gaussian window function specifying the field-detector interaction, only enters as an overall factor in the negativity and hence plays no role in determining its sign (and thus whether entanglement has been transfered to the detectors). In this letter, we go beyond the stationary phase approximation to compute the non-trivial $c_3$ dependence.

\section{Basic Set-Up}\label{set-up}
The simplest setting to study entanglement harvesting is that of two accelerating Unruh-DeWitt detectors labeled $A$ and $B$, each described by a two-state Hamiltonian \(H_{\textrm{det}_i}\) of the form

\[H_{\textrm{det}_i} = \frac{\Omega}{2} \left(\left\vert \uparrow \right> \left< \uparrow \right\vert - \left\vert \downarrow \right> \left< \downarrow \right\vert\right)\]
acting on the detector Hilbert spaces \(\mathcal{H}_{\textrm{det}_i}\) and each coupled to the same scalar field $\phi$ through an interaction Hamiltonian

\[H_{\textrm{int}_i} = \eta(\tau) \phi(x_i(\tau)) \left( \left\vert \uparrow \right> \left< \downarrow \right\vert + \left\vert \downarrow \right> \left< \uparrow \right\vert \right) \]
where $x_i(\tau)$ parameterizes the worldline of detector $i$ ($i = A, B$) in terms of the detector's proper time $\tau$. 
We envision that the Unruh-DeWitt detectors are travelling along
worldlines with constant acceleration either parallel or anti-parallel
to one another. In such a model, the overall Hilbert space is
\(\mathcal{H}_{\phi} \otimes \mathcal{H}_{\textrm{det}_A} \otimes \mathcal{H}_{\textrm{det}_B}\)
and the Hamiltonian is given by

\[H = H_{\phi} + H_{\textrm{det}_A} + H_{\textrm{det}_B} + H_{\textrm{int}_A} + H_{\textrm{int}_B}\]
where \(H_{\textrm{det}_i}\) and \(H_{\textrm{int}_i}\) represent the
internal Hamiltonians of the detectors and interaction Hamiltonians
respectively. Switching to the interaction picture, we find that the
interaction Hamiltonian is again the sum of the Hamiltonians for each
individual detector. We calculate the final state of the system at
\(\tau = \infty\) after starting in the state
\(\left\vert 0 \right> \left\vert \downarrow_A \right> \left\vert \downarrow_B \right>\)
at \(\tau = - \infty\). Define the operators \(\Phi^{\pm}_{i}\) for
\(i = A, B\) by

\[ \Phi^{\pm}_i = \int_{-\infty}^{\infty}d\tau' \eta(\tau') e^{\pm i \Omega \tau'} \phi(x_{i}(\tau')) \]
Now define an operator \(S\) by
\[S = -i \sum_{i=A,B} \sum_{j=+,-} \Phi^{j}_{k} \sigma^{j}_{k} \] where
\(\sigma^{+}_{k} = \left\vert \uparrow_k \right> \left< \downarrow_k \right\vert\)
and
\(\sigma^{-}_{k} = \left\vert \downarrow_k \right> \left< \uparrow_k \right\vert\)
Note that

\[ S^2 = - \left( \Phi^-_A \Phi^+_A  + \Phi^-_B \Phi^+_B + \Phi^+_B \Phi^+_A \left\vert \uparrow_A \right> \left\vert \uparrow_B \right> \left< \downarrow_A \right\vert \left< \downarrow_B \right\vert + \Phi^+_A \Phi^+_B \left\vert \uparrow_A \right> \left\vert \uparrow_B \right> \left< \downarrow_A \right\vert \left< \downarrow_B \right\vert \right) +\textrm{(irrelevant terms)}\]
where "irrelevant terms" refers to terms which vanish when considering the action of $S^2$ on the ground state of the system.  

The time evolution operator in the interaction picture is then given by
\(\mathcal{T}\left[e^{S}\right]\), so that to second order in
perturbation theory the state \(\left\vert \psi \right>\) at \(\tau = \infty\) is given by \[
\begin{aligned}
\left\vert \psi \right> = \left(1 + S + \frac{1}{2} \mathcal{T}[SS]\right) \left\vert 0 \right> \left\vert \downarrow_A \right> \left\vert \downarrow_B \right> & = & (1 + d_1) \left\vert 0 \right> \left\vert \downarrow_A \right> \left\vert \downarrow_B \right> \\ 
& & - i \left( \Phi^{+}_A \left\vert 0 \right> \left\vert \uparrow_A \right> \left\vert \downarrow_B \right> + \Phi^{+}_B \left\vert 0 \right> \left\vert \downarrow_A \right> \left\vert \uparrow_B \right> \right) \\ 
& & -\frac{1}{2} \mathcal{T}\left[ \Phi^+_A \Phi^+_B + \Phi^+_B \Phi^+_A \right] \left\vert 0 \right> \left\vert \uparrow_A \right> \left\vert \uparrow_B \right> \\
& & + \left( \frac{1}{2} \mathcal{T}\left[ \Phi^-_A \Phi^+_A + \Phi^-_B \Phi^+_B \right] \left\vert 0 \right> - d_1 \left\vert 0 \right> \right) \left\vert \downarrow_A \right> \left\vert \downarrow_B \right>
\end{aligned}
\] where we have defined
\(d_1 = -\frac{1}{2} \left< 0 \vert \mathcal{T}\left[ \Phi^-_A \Phi^+_A + \Phi^-_B \Phi^+_B \right] \vert 0 \right>\).
The expression on the last line contains a field state factor which is orthogonal to the field ground state $\left\vert 0 \right>$. 

We find the density matrix corresponding to this pure state to second
order in perturbation theory. Keeping only terms with at most two
\(\Phi^k_j\) factors in them and which are nonvanishing after taking
partial trace over \(\mathcal{H}_{\phi}\), we obtain \[
\begin{aligned}
\rho = \left\vert \psi \right> \left< \psi \right\vert & = & (1 + d_1 + d^*_1) \left\vert 0 \right> \left\vert \downarrow_A \right> \left\vert \downarrow_B \right> \left< 0 \right\vert \left< \downarrow_A \right\vert \left< \downarrow_B \right\vert \\ & &
- \Phi^+_A \Phi^+_B \left\vert 0 \right> \left\vert \uparrow_A \right> \left\vert \uparrow_B \right> \left< 0 \right\vert \left< \downarrow_A \right\vert \left< \downarrow_B \right\vert \\ & &
-  \left\vert 0 \right> \left\vert \downarrow_A \right> \left\vert \downarrow_B \right> \left( \left< 0 \right\vert \Phi^-_B \Phi^-_A \right) \left< \uparrow_A \right\vert \left< \uparrow_B \right\vert \\ & &
- \Phi^+_A \left\vert 0 \right> \left\vert \uparrow_A \right> \left\vert \downarrow_B \right> \left( \left< 0 \right\vert \Phi^{-}_A \right) \left< \uparrow_A \right\vert \left< \downarrow_B \right\vert \\ & &
- \Phi^+_B \left\vert 0 \right> \left\vert \downarrow_A \right> \left\vert \uparrow_B \right> \left( \left< 0 \right\vert \Phi^{-}_B \right) \left< \downarrow_A \right\vert \left< \uparrow_B \right\vert \\ & &
- \Phi^+_A \left\vert 0 \right> \left\vert \uparrow_A \right> \left\vert \downarrow_B \right> \left( \left< 0 \right\vert \Phi^{-}_B \right) \left< \downarrow_A \right\vert \left< \uparrow_B \right\vert \\ & &
- \Phi^+_B \left\vert 0 \right> \left\vert \downarrow_A \right> \left\vert \uparrow_B \right> \left( \left< 0 \right\vert \Phi^{-}_A \right) \left< \uparrow_A \right\vert \left< \downarrow_B \right\vert
\end{aligned}
\] which, after partial tracing, becomes \[
\begin{aligned}
\rho_{\textrm{tr}} = \left\vert \psi \right> \left< \psi \right\vert & = & (1 + d_1 + d^*_1) \left\vert \downarrow_A \right> \left\vert \downarrow_B \right> \left< \downarrow_A \right\vert \left< \downarrow_B \right\vert \\ & &
- \left< 0 \vert \Phi^+_A \Phi^+_B \vert 0 \right> \left\vert \uparrow_A \right> \left\vert \uparrow_B \right> \left< \downarrow_A \right\vert \left< \downarrow_B \right\vert \\ & &
- \left< 0 \vert \Phi^+_A \Phi^+_B \vert 0 \right>^* \left\vert \downarrow_A \right> \left\vert \downarrow_B \right> \left< \uparrow_A \right\vert \left< \uparrow_B \right\vert \\ & &
- \left< 0 \vert \Phi^-_A \Phi^+_A \vert 0 \right> \left\vert \uparrow_A \right> \left\vert \downarrow_B \right> \left< \uparrow_A \right\vert \left< \downarrow_B \right\vert \\ & &
-\left< 0 \vert \Phi^-_B \Phi^+_B \vert 0 \right> \left\vert \downarrow_A \right> \left\vert \uparrow_B \right> \left< \downarrow_A \right\vert \left< \uparrow_B \right\vert \\ & &
- \left< 0 \vert \Phi^-_B \Phi^+_A \vert 0 \right> \left\vert \uparrow_A \right> \left\vert \downarrow_B \right> \left< \downarrow_A \right\vert \left< \uparrow_B \right\vert \\ & &
- \left< 0 \vert \Phi^+_B \Phi^-_A \vert 0 \right> \left\vert \downarrow_A \right> \left\vert \uparrow_B \right> \left< \uparrow_A \right\vert \left< \downarrow_B \right\vert
\end{aligned}
\]

Writing this in matrix form, we find

\[
\rho_{\textrm{tr}} = \left(\begin{array}{cccc}
1 - \left< 0 \vert \Phi^-_A \Phi^+_A + \Phi^-_B \Phi^+_B \vert 0 \right> & - \left< 0 \vert \Phi^+_A \Phi^+_B \vert 0 \right> & 0 & 0 \\
- \left< 0 \vert \Phi^+_A \Phi^+_B \vert 0 \right> & 0 & 0 & 0 \\
0 & 0 & \left< 0 \vert \Phi^-_A \Phi^+_A \vert 0\right> & \left< 0 \vert \Phi^-_B \Phi^+_A \vert 0\right>  \\
0 & 0 & \left< 0 \vert \Phi^-_A \Phi^+_B \vert 0\right>  & \left< 0 \vert \Phi^-_B \Phi^+_B \vert 0\right> \\
\end{array}\right)
\]

We define \[
E_A =  \left< 0 \vert \Phi^-_A \Phi^+_A \vert 0 \right> = \int_{-\infty}^{\infty} d\tau' \int_{-\infty}^{\infty} d\tau'' \eta(\tau') \eta(\tau'') e^{i \Omega (\tau' - \tau'')} G(x_A(\tau'), x_A(\tau''))
\]

\[
E_B = \left< 0 \vert \Phi^-_B \Phi^+_B \vert 0 \right> = \int_{-\infty}^{\infty} d\tau' \int_{-\infty}^{\infty} d\tau'' \eta(\tau') \eta(\tau'') e^{i \Omega (\tau' - \tau'')} G(x_B(\tau'), x_B(\tau''))
\]

\[
X = \left< 0 \vert \Phi^+_A \Phi^+_B \vert 0 \right> = \int_{-\infty}^{\infty} d\tau' \int_{-\infty}^{\tau'} d\tau'' \eta(\tau') \eta(\tau'') e^{i \Omega (\tau' + \tau'')} G(x_A(\tau'), x_B(\tau''))
\]
where \(G\) is the Feynman propagator for \(\phi\) and we have made use of the symmetry under exchanging $\tau' \leftrightarrow \tau''$ to rewrite the $E_i$ integrals as being over the entire $\tau' - \tau''$ plane.

\subsection{Parallel Worldlines}\label{parallel-worldlines}

Salton et al. investigate such a situation. A massless field \(\phi\) is
coupled to two detectors with worldlines denoted by \(x_A(\tau)\) and
\(x_B(\tau)\). In the parallel case, the detector worldlines are of the
form

\[x_A(\tau) = (t = \frac{1}{\kappa} \sinh{\kappa \tau}, x = \frac{1}{\kappa}\left(\cosh{\kappa \tau} - 1 \right), y = 0, z = 0) \]
\[x_B(\tau) = (t = \frac{1}{\kappa} \sinh{\kappa \tau}, x = \frac{1}{\kappa}\left(\cosh{\kappa \tau} - 1 \right) + L, y = 0, z = 0) \]

and the Feynman propagator for a massless field \(\phi\) is given by

\[G(x,y) = -\frac{1}{(2 \pi)^2} \frac{1}{(x-y)^2}\]

These choices lead to the integrals (in this case, \(E_A = E_B = E\))

\[ 
\begin{aligned}
E & = & -\frac{\kappa^2}{4 \pi^2} \int_{-\infty}^{\infty} d\tau' \int_{-\infty}^{\infty} d\tau'' \exp{\left[ - \frac{1}{2 \sigma^2} (\tau'^2 + \tau''^2) - i \Omega (\tau' - \tau'') \right]} \\ & & \times \frac{1}{(\sinh{\kappa \tau'} - \sinh{\kappa \tau''})^2 - (\cosh{\kappa \tau'} - \cosh{\kappa \tau''})^2}
\end{aligned}
\]

\[
\begin{aligned}
X & = & -\frac{\kappa^2}{4 \pi^2} \int_{-\infty}^{\infty} d\tau' \int_{-\infty}^{\tau'} d\tau'' \exp{\left(-\frac{\tau'^2+\tau''^2}{2 \sigma^2} + i \Omega (\tau' + \tau'')\right)} \\ & & \times \frac{1}{(\sinh{\kappa \tau'} - \sinh{\kappa \tau''})^2 - (\cosh{\kappa \tau'} - \cosh{\kappa \tau''} - L \kappa)^2}
\end{aligned}
\]

We first note that if we define \(x = \tau' + \tau''\) and
\(y = \tau' - \tau''\) after some algebraic manipulation, the integrals
can be rewritten as \[ 
\begin{aligned}
E & = & -\frac{\kappa^2 \eta^2_0}{32 \pi^2} e^{-\sigma^2 \Omega^2} \int_{-\infty}^{\infty} dx \int_{-\infty}^{\infty} dy \exp{\left[ - \frac{1}{4 \sigma^2} (x^2 + (y + 2 i \Omega \sigma^2)^2) \right]} \textrm{csch}^2\left(\frac{\kappa y}{2} \right)
\end{aligned}
\]

\[
\begin{aligned}
X & = & \frac{\kappa^2 \eta^{2}_{0}}{32 \pi^2} e^{-\sigma^2 \Omega^2} \int_{-\infty}^{\infty} dx \int_{0}^{\infty} dy \exp{\left(-\frac{(x-2 i \Omega \sigma^2)^2+y^2}{4 \sigma^2} \right)} \\ & & \times \left[ \frac{L \kappa}{2} + i \epsilon - e^{-\kappa x /2} \text{sinh}\left(\frac{\kappa y}{2}\right) \right]^{-1} \left[ \frac{L \kappa}{2} - i \epsilon + e^{\kappa x /2} \text{sinh}\left(\frac{\kappa y}{2}\right) \right]^{-1} 
\end{aligned}
\]

Introducing the dimensionless parameters \(c_1 = \kappa L\),
\(c_2 = \kappa \Omega \sigma^2\), and \(c_3 = \sigma \Omega\) and
substituting \(\tilde{x} = x/L\) and \(\tilde{y} = y/L\), we obtain

\[ 
\begin{aligned}
E & = & -\eta^2_0 \frac{c_1^2}{32 \pi^2} e^{-c_3^2} \int_{-\infty}^{\infty} d\tilde{x} \int_{-\infty}^{\infty} d\tilde{y} \exp{\left[ - \left(\frac{c_1 c_3}{2 c_2}\right)^2 \left(\tilde{x}^2 + (\tilde{y} + 2 i c_2/c_1)^2\right) \right]} \textrm{csch}^2\left(\frac{c_1 \tilde{y}}{2} \right)
\end{aligned}
\]

\[
\begin{aligned}
X & = & \eta^{2}_{0} \frac{c_1^2}{32 \pi^2} e^{-c_3^2} \int_{-\infty}^{\infty} d\tilde{x} \int_{0}^{\infty} d\tilde{y} \exp{\left(-\left(\frac{c_1 c_3}{2 c_2}\right)^2 ((\tilde{x}-2 i c_2 / c_1)^2+\tilde{y}^2) \right)} \\ & & \times \left[ \frac{c_1}{2} + i \epsilon - e^{-c_1 \tilde{x} /2} \text{sinh}\left(\frac{c_1 \tilde{y}} {2}\right) \right]^{-1} \left[ \frac{c_1}{2} - i \epsilon + e^{c_1 x /2} \text{sinh}\left(\frac{c_1 y}{2}\right) \right]^{-1} 
\end{aligned}
\]

We can make these integrals easier to evaluate by shifting the contour
of integration in the complex plane. In the case of \(E\), we shift from
integrating \(\tilde{y}\) along the real axis to integrating along the
line \(y = y' - 2 i c_2 / c_1\) where \(y'\) ranges from \(-\infty\) to
\(\infty\). In the case of \(X\), we shift from integrating \(x\) along
the real axis to integrating along the line
\(\tilde{x} = x' + 2 i c_2 / c_1\) where \(y'\) ranges from \(-\infty\)
to \(\infty\). Note that this allows us to neglect the \(i \epsilon\) in
our denominators, since the integrals are no longer crossing poles. The
resulting integrals are

\[ 
\begin{aligned}
E & = & -\eta^2_0 \frac{c_1^2}{32 \pi^2} e^{-c_3^2} \int_{-\infty}^{\infty} d\tilde{x} \int_{-\infty}^{\infty} d\tilde{y} \exp{\left[ - \left(\frac{c_1 c_3}{2 c_2}\right)^2 \left(\tilde{x}^2 + \tilde{y}^2\right) \right]} \textrm{csch}^2\left(\frac{c_1 \tilde{y}}{2} - i c_2 \right)
\end{aligned}
\]

\[
\begin{aligned}
X & = & \eta^{2}_{0} \frac{c_1^2}{32 \pi^2} e^{-c_3^2} \int_{-\infty}^{\infty} d\tilde{x} \int_{0}^{\infty} d\tilde{y} \exp{\left(-\left(\frac{c_1 c_3}{2 c_2}\right)^2 
(\tilde{x}^2+\tilde{y}^2) \right)} \\ & & \times \left[ \frac{c_1}{2} + i \epsilon - e^{-c_1 \tilde{x} /2} e^{-i c_2} \text{sinh}\left(\frac{c_1 \tilde{y}} {2}\right) \right]^{-1} \left[ \frac{c_1}{2} - i \epsilon + e^{c_1 x /2} e^{i c_2} \text{sinh}\left(\frac{c_1 y}{2}\right) \right]^{-1} 
\end{aligned}
\]

The expression for \(E\) can be further simplified by noting that the
integral over \(\tilde{x}\) is purely Gaussian. Carrying out the
\(\tilde{x}\)-integral yields

\[ 
\begin{aligned}
E & = & -\eta^2_0 \frac{c_1}{16 \pi^{3/2}} \frac{c_2}{c_3} e^{-c_3^2} \int_{-\infty}^{\infty} d\tilde{y} \exp{\left[ - \left(\frac{c_1 c_3}{2 c_2}\right)^2 \tilde{y}^2 \right]} \textrm{csch}^2\left(\frac{c_1 \tilde{y}}{2} - i c_2 \right)
\end{aligned}
\]

The paper by Salton et al. uses the stationary phase approximation on
both of these integrals. Given our shift of variables, this is
equivalent to replacing the factor \(f(x,y)\) multiplying the Gaussian
in each by \(f(0,0)\). This gives

\[ 
\begin{aligned}
E_{\textrm{sp}} & = & \eta^2_0 \frac{e^{-c_3^2}}{8 \pi} \left(\frac{c_2}{c_3}\right)^2  \csc^2{c_2}
\end{aligned}
\]

\[
\begin{aligned}
X_{\textrm{sp}} & = & \eta^{2}_{0} \frac{e^{-c_3^2}}{2 \pi} \left(\frac{c_2}{c_3 c_1}\right)^2
\end{aligned}
\]
for the integrals.  Whether the detectors are entangled is determined by calculating whether the negativity $\mathcal{N}$ of the system described by $\rho_\textrm{tr}$ is non-zero.  The negativity, discussed as means for measuring entanglement in \cite{Peres} \cite{Horodecki}, is given in this situation by

\[
\mathcal{N} = \max\{\vert X \vert - E,0\} = \eta_0^2 \frac{e^{-c_3^2}}{8 \pi} \left(\frac{c_2}{c_3}\right)^2 \left[ \frac{4}{c_1^2} -  \csc^2{c_2} \right]
\]
In this approximation $c_3$ only enters in an overall factor and so has no impact on the sign of $\mathcal{N}$.

For large values of $c_3$, the Gaussian factor in the integrands for both $E$ and $X$ suppresses the integrand everywhere except for the point $(\tilde{x} = 0, \tilde{y} = 0)$.  This suggests that for large $c_3$, we should obtain the same result as we would obtain using the stationary phase approximation.  Physically, for fixed $c_1$, $c_2$ we find that $c_3$ parameterizes the width of the window function Gaussian and so the amount of time the detectors have to interact with each other.  This leads to the expectation that for small $c_3$ there will not be enough time for entanglement to be established, while for large $c_3$ the presence of entanglement is dependent on the parameters governing the choice of detector worldlines.

\section{Numerical Evaluation}\label{numerical-evaluation}

We numerically evaluated the integrals for \(E\) and
\(X\) for the range of parameter space \(c_1 \in [0,6]\),
\(c_2 \in [0,3]\), \(c_3 \in [0,5]\), using Mathematica. For \(E\), the single integral
can be evaluated straightforwardly using a Gauss-Kronrod method (with is Mathematica's default
one-dimensional numerical integration algorithm). For \(X\), using the default ``GlobalAdaptive''
strategy with a multidimensional Gauss-Kronrod method, Mathematica warns
about possible inaccuracy when evaluating the integral for some
parameters within the chosen parameter space. Although Mathematica
reports a guess for the error on these numerical integrals, there is no
guarantee that the guess will not greatly underestimate the true amount
of error. We provide our own estimate of the amount of error by doing
the integrations using Mathematica's ``LocalAdaptive'' strategy rather
than its default ``GlobalAdaptive'' strategy. Both strategies in this
case compute numerical integrals by recursively dividing up the
integration region into subregions and using a Gauss-Kronrod method to
estimate the integral value and error. However, the ``LocalAdaptive''
strategy makes its choice of which subregion to further divide via a
local estimate of the integration error in that region, while
``GlobalAdaptive'' chooses which subregions to refine based on the
magnitude of error compared to the overall value of the integral.

We calculated the values of \(E\) and \(X\) on the 3D grid in parameter
space on which the parameters take on the values
\(c_1 = \{0.025, 0.050, 0.075, ..., 5.975, 6.000\}\),
\(c_2 = \{0.025, 0.050, 0.075, ..., 3.000\}\),
\(c_3 = \{0.125, 0.250, ..., 4.875, 5.000\}\) using the
``LocalAdaptive'' integration strategy. We also calculated
\(X\) using the ``GlobalAdaptive'' strategy on the same grid to compare
with the ``LocalAdaptive'' results.

We then used the values of \(E\) and \(X\) to calculate
\(\mathcal{N} = \vert X \vert - E\), and used the sign of
\(\mathcal{N}\) to determine which regions of parameter space support
entanglement. The regions are shown in 5 figures.

We calculated the difference between the ``LocalAdaptive'' and
``GlobalAdaptive'' results for both the values of \(E\) and \(X\) as
well as the final negativity result \(\mathcal{N}\). We found that the
values for \(X\) matched to within \(0.08\%\) and for \(\mathcal{N}\) to
within \(9\%\). \\
\begin{figure}
\includegraphics[scale=0.5]{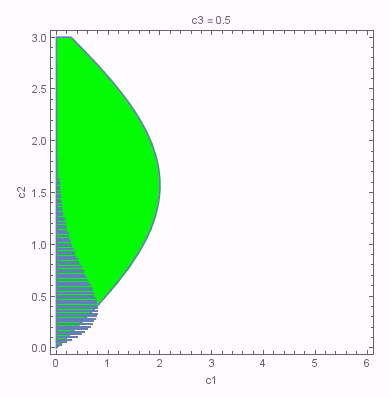} 
\includegraphics[scale=0.5]{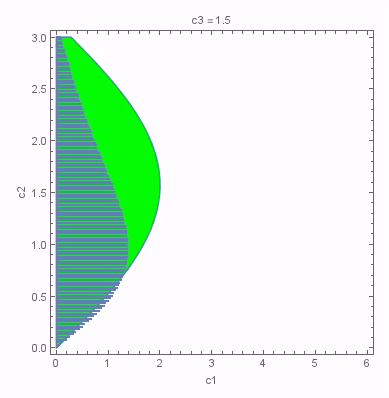} \\
\includegraphics[scale=0.5]{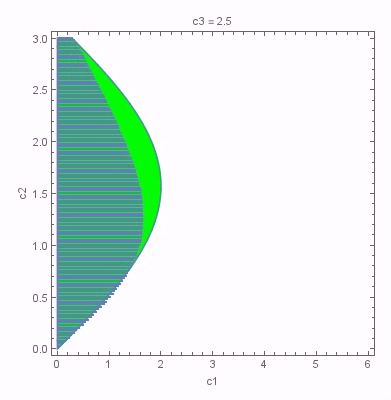} 
\includegraphics[scale=0.5]{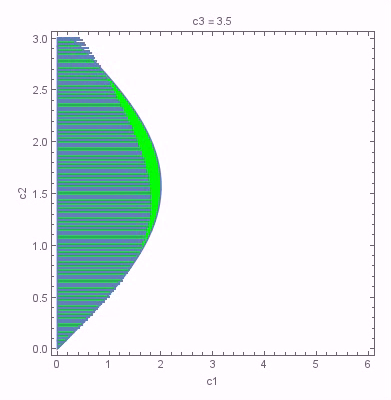} \\
\includegraphics[scale=0.5]{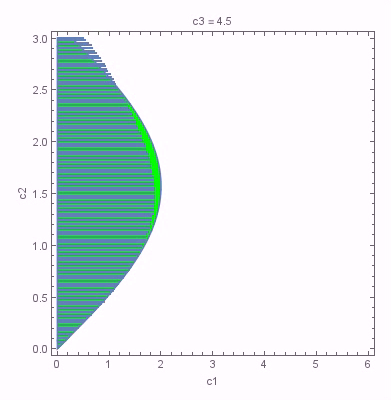}
\caption{Plots of the entanglement region for $c_3 = 0.5, 1.5, 2.5, 3.5, 4.5$.  The green region is the entanglement region in the stationary phase approximation, while the blue overlay denotes points in parameter space for which entanglement was established numerically.}
\end{figure}
Collectively, this allowed us to determine the dependence of the entanglement region on \(c_3\).  In the limit as \(c_3\) approaches \(0\), the entanglement region vanishes, while for \(c_3 \geq 4.5\) the entanglement region looks similar to that computed by Salton et al.  This is consistent with the expectation that the stationary phase approximation integrals for \(E\) and \(X\) should be accurate for large \(c_3\).

\section{Discussion}

A natural next step in this line of research is to extend the analysis to more general trajectories, including the antiparallel case, and, of significant interest to a complete analysis, to consider the effect of different window functions. The latter would also us, for example, to establish that the Gaussian tails of the window functions currently in use play no essential role in the entanglement results. We intend to return to these  undertakings in future work.



\begin{thebibliography}{8}
\bibitem{Salton}Salton, Grant, Robert B. Mann, and Nicolas C. Menicucci. "Acceleration-assisted entanglement harvesting and rangefinding." New J. Phys. 17.3 (2015): 035001.
\bibitem{Nambu}Nambu, Yasusada. "Entanglement structure in expanding universes." Entropy 15.5 (2013): 1847-1874.
\bibitem{Martin-Martinez}Martin-Martinez, Eduardo, and Nicolas C. Menicucci. "Cosmological quantum entanglement." Class. Quantum Grav. 29.22 (2012): 224003.
\bibitem{VerSteeg}Ver Steeg, Greg, and Nicolas C. Menicucci. "Entangling power of an expanding universe." Phys. Rev. D 79.4 (2009): 044027.
\bibitem{Reznik}Reznik, Benni, Alex Retzker, and Jonathan Silman. "Violating Bell's inequalities in vacuum." Phys. Rev. A 71.4 (2005): 042104.
\bibitem{Reznik2}Reznik, Benni. "Entanglement from the vacuum." Found. Phys. 33.1 (2003): 167-176.
\bibitem{Peres}A. Peres, Phys. Rev. Lett. 77, 1413 (1996)
\bibitem{Horodecki}M. Horodecki, P. Horodecki and R. Horodecki, Phys. Lett. A 223, 1 (1996)
\end{thebibliography}
\end{document}